\documentstyle{amsppt}
\TagsOnRight
\catcode`\@=11
\def\logo@{}
\catcode`\@=13
\parindent=8 mm
\magnification 1200
\hsize = 12.2  cm
\vsize = 19.4  cm
\hoffset = 1 cm
\parskip=\medskipamount
\baselineskip=14pt
\rightline{ CRM-2338 (1996) \break}
\rightline{solv-int/9603001 \break} 

\def \tr {\text{tr}}
\def \diag {\text{diag}}

\def \smaller {\eightpoint}
\def \wt {\widetilde}

\def \ra {\rightarrow}

\def \a {\alpha}
\def \b {\beta}

\def \g {\gamma}

\def \l {\lambda}
\def \L {\Lambda}

\def \z {\zeta}

\def \KK {\Cal K}
\def \LL {\Cal L}

\def \NN {\Cal N}
\def \LL {\Cal L}
\def \OO {\Cal O}

\def \SS {\Cal S}

\def \di {\partial}

\hyphenation{Darboux Liouville Hurtubise Harnad Adams Arnold Helminck}
\font\titlefont=cmbx10 scaled\magstep2
\font\sectionfont=cmbx10 at 14pt

\vskip 1 cm
\noindent{\titlefont
Darboux Coordinates on Coadjoint Orbits of \newline \noindent Lie Algebras}
${}^\dag$
\rightheadtext{}
\leftheadtext{}
\footnote""
{\hskip -.8 cm ${}^\dag$
Research supported in part by the Natural Sciences and Engineering Research
 Council of Canada, the Fonds FCAR du Qu\'ebec and National Science Foundation
grants DMS-9106938 and DMS-9305857. One of the authors would like to thank the 
University of Kansas for their hospitality during the writing of this paper.}
\bigskip \bigskip \noindent M.R. ~Adams${}^1$, J.~Harnad${}^2$, J.
~Hurtubise${}^3$
\medskip
\noindent{\eightpoint $^1$ Department of Mathematics \newline \noindent
University of Georgia \newline \noindent Athens, GA 30602 USA \newline
\noindent e-mail: adams\@alpha.math.uga.edu   
\medskip 
\noindent   $^2$
Department of Mathematics and Statistics \newline \noindent Concordia University
\newline \noindent
 7141 Sherbrooke W., Montr\'eal, Canada H4B 1R6,\newline
\noindent and 
\newline
\noindent
 Centre de recherches math\'ematiques \newline \noindent Universit\'e de
Montr\'eal \newline \noindent
  C.~P. 6128-A, Montr\'eal, Canada H3C 3J7 \newline \noindent e-mail:  
harnad\@alcor.concordia.ca  {\it \ or \ } 
 harnad\@crm.umontreal.ca 
\medskip 
\noindent   $^3$
Department of Mathematics 
\newline \noindent McGill University
\newline \noindent
 Montr\'eal, Canada H3A 2K6 \newline \noindent
e-mail: hurtubis\@gauss.math.mcgill.ca
\medskip}
\bigskip \bigskip

\noindent {\bf Abstract.} The method of 
constructing spectral Darboux coordinates on finite dimensional  coadjoint
orbits in
duals of loop algebras is applied  to the one pole case, where the orbit is
identified with a coadjoint orbit in the dual of a finite dimensional Lie
algebra.  The constructions are carried out explicitly when the Lie algebra is
$\frak{sl}(2,
\bold R),\ 
\frak{sl}(3, \bold R),$ and $\frak{so}(3, \bold R)$, and for rank two 
orbits in $\frak{so}(n, \bold
R)$.  A new feature that appears is the possibility of identifying spectral
Darboux coordinates
associated to ``dynamical" choices of sections of the associated 
eigenvector line bundles; i.e.
sections that depend on the point within the given orbit.  
Darboux coordinates.  
\bigskip \newpage \noindent
{\sectionfont 1 Introduction} \medskip

Many finite dimensional integrable Hamiltonian systems have been realized
through 
symplectic reduction {\bf [AM]} or
collectivization {\bf [GS]}  from integrable Hamiltonian systems on coadjoint
orbits in the dual of Lie algebras.  The classic example of this is the motion
of  a free rigid body, whose natural phase space   $T^*SO(3)$  reduces to
$\frak{so}(3)^*$ after
factoring  out the symmetries given by the left
$SO(3)$ action.  The reduced dynamics are given by the Euler equations on
$\frak{so}(3)^*$
{\bf [Ar]}.  Other well-known examples include the symmetric tops (i.e.
rigid bodies in external
force fields),  constrained oscillators on quadrics, and Toda
lattices.   In view of such examples it is interesting to make a general
study of integrable
Hamiltonian systems on coadjoint orbits.   (This is also interesting from
the geometric quantization
point of view since  a completely integrable system yields a -possibly
singular- real polarization
of the coadjoint orbit.)

There  are two widely used methods  for producing completely integrable
systems on coadjoint
orbits:  the Mischenko, Fomenko technique  {\bf [MF]}, and the Adler,
Kostant, Symes  (AKS) theorem
{\bf [Ad], [K], [S]}.  In fact, the former can be understood as a special
case of the latter.   It
 relies on the observation that if two functions $F_1, F_2$ Poisson
commute on
$\frak{g}^*$, and
$\mu_0 \in
\frak{g}^*$ then the family of shifted functions $F_i^{\lambda}$ defined by
$F_i^{\lambda}(\nu) = F_i(\lambda \mu_0 +\nu)$ Poisson commute for all scalars
$\lambda$.   The AKS theorem gives Poisson commuting functions on the dual of a
Lie algebra $\frak{k}^*$ when $\frak{g}$ splits into a vector space direct
sum $\frak{g} =
\frak{k} \oplus \frak{l}$ with both $\frak{k}$ and $\frak{l}$ Lie subalgebras. 
It  states that the $Ad^*$ - invariant functions on
$\frak{g}^*$ then restrict to Poisson commuting functions on $\frak{k}^*$. 
 More generally, if the
$Ad^*$ - invariant functions are shifted by an infinitesimal character of
$\frak{l}$ (in $\frak
{l}^*$), one still obtains a ring of Poisson commuting functions on
$\frak{k}^*$.  To treat periodic
Toda lattices and constrained oscillators in such a framework requires use
of coadjoint orbits in
loop algebras.  This places such systems in the same context as the
invariant ``finite dimensional"
solutions to integrable systems of PDE's such as the KdV equation.

  Most of the above mentioned families of
integrable systems can be studied as shifted AKS type flows on finite
dimensional orbits of loop algebras (see e.g. {\bf [AvM], [R], [AHP], [AHH3]}). 
In particular if $\frak{g}$ denotes a Lie algebra of matrices we let
$\wt{\frak{g}}$ denote the formal loop algebra
$$
\wt{\frak{g}} = \{\sum_{i = -\infty}^n X_i\lambda^i \ |\ X_i \in \frak{g}\}
\tag{1.1}
$$
and split $\wt{\frak{g}} = \wt{\frak{g}}^+ \oplus \wt{\frak{g}}_-$ where
$\wt{\frak{g}}^+$ consists of those loops that are polynomial in $\lambda$
(i.e. matricial polynomials) and $\wt{\frak{g}}_-$ consists of  loops with
only strictly negative powers of $\lambda$.  If $\frak{g}$ has a nondegenerate
bilinear form $< , >$ we may use it to give an identification
$$ \frak{g} \sim \frak{g}^* . \tag{1.2}
$$
We can extend $< , >$ to $\wt{\frak{g}}$ by computing $<X(\lambda),Y(\lambda)>$
as a formal power series  and then taking the coefficient of $\lambda^{-1}$. 
This gives an identification
$$
(\wt{\frak{g}}^+)^* \sim \wt{\frak{g}}_- . \tag{1.3}
$$
If $< , >$ is $Ad$ invariant one can compute the $ad^*$ action of
$\wt{\frak{g}}^+$ on $\wt{\frak{g}}_-$ and thus determine the symplectic leaves
in $\wt{\frak{g}}_-$ with respect to the Poisson structure of
$(\wt{\frak{g}}^+)^*$.  In
particular, there is a Poisson embedding of the finite dimensional space,
$\frak{g}^*$ into
$(\wt{\frak{g}}^+)^*$ which under the identifications (1.2) and (1.3) is simply
$$
X \rightarrow \frac{1}{\lambda}X  \tag{1.4}
$$ for $X \in \frak{g}$.
In the next section we explain how 
Mischenko - Fomenko type Hamiltonians on $\frak{g}^*$ can be realized as
shifted AKS Hamiltonians on
$(\wt{\frak{g}}^+)^*$ by means of the map (1.4).  Generalizing to twisted
loop algebras, AKS type
Hamiltonians on
$\frak{g}^*$ can also be derived in this manner.

   In {\bf [AHH2]} the shifted AKS flows were studied on finite dimensional
orbits in $(\wt{\frak{g}}^+)^*$ through points of the form 
$$
\NN_0(\l) = \sum_{i=1}^n 
{N_i \over \l - \a_i},  \tag{1.5}
$$
where $N_i \in \frak{g}$ and the $\a_i$'s  are scalar.  It was shown how the
spectral data on these orbits led directly to algebraic Darboux coordinates
that are
particularly suited to integrating the flows of shifted AKS type
Hamiltonians.  Since these
coordinates are obtained from a divisor of a section of the dual
eigenvector line bundle over the
invariant spectral curve defined by the characteristic equation of
$\NN_0(\l)$, we call them {\it
spectral Darboux coordinates}.  In section 3 of this  letter we  show how  
to obtain spectral
Darboux coordinates on coadjoint orbits of finite dimensional Lie algebras
and we
illustrate the construction explicitly for
$\frak{g} =
\frak{sl}(2),
\frak{sl}(3)$ and $\frak{so}(3).$  Since the Euler equations for rigid body
motion are Mischenko -
Fomenko type flows on $\frak{so}(3)^*$, this case produces Darboux
coordinates particularly
suited to integrating rigid body flow.  More generally, these techniques
may be used to study
the
$\frak{so}(n)$  generalization of rigid body motion.  Unfortunately, as the
size of the matrix
increases  so do  the the degrees of the polynomials whose roots describe
the Darboux coordinates,
so these coordinates are no longer computable for large n.  On the other
hand, it was pointed
out by Moser {\bf[Mo]} that certain rigid body flows in $\frak{so}(n)^*$
(i.e. flows on very
special coadjoint orbits) can be dealt with explicitly in terms of rank two
perturbations of an
$n\times n$ matrix.  In {\bf [AHH3]} it was shown that this is an example
of a general phenomenon,
referred to as {\it duality}, relating spectral curves of two matricial
polynomials of different
size.  In section 4 of this letter we describe the Darboux coordinates on
this special orbit for the
$\frak{so}(n)$ rigid body and compare them with those obtained through duality.
 
\bigskip
\noindent{\sectionfont 2  Mischenko Fomenko Flows as Shifted  \newline AKS
Flows}
\medskip

     For convenience we take $\frak{g} = \frak{sl}(r)$ and $G = SL(r)$. 
The $Ad^*$ - invariant
inner product can then be taken to be
$$
<X,Y> = tr(XY^T)    \tag{2.1}
$$ 
 and the $Ad^*$ - invariant functions are generated by 
$$
\sigma_k(X) := \frac{1}{k} tr(X^k). \tag{2.2}
$$
Fixing $Y \in \frak{g}$ and $\mu \in \Bbb R$, Hamilton's equation for the
Mischenko - Fomenko 
Hamiltonian
$\sigma_k(X + \mu Y)$ is
$$
\frac{d}{dt} X = [(X + \mu Y)^{k-1},X].  \tag{2.3}
$$
On the other hand, the $Ad^*$ - invariant functions on $\wt{\frak{g}}^*$   (identified with
$\wt{\frak{g}}$ by (1.3)) are generated by
$$
S_{k,j}(X(\lambda)) := \frac{1}{k}tr(X(\lambda)^k))_j  \tag{2.4}
$$
where the subscript $j$ denotes taking the coefficient of $\lambda^j$. 
Shifting $S_{k,-j}$ by   
$Y \in \frak{g}$ and restricting to the union of orbits $\{X(\lambda) =
\frac{1}{\lambda}X\}
\subset
\wt{\frak{g}}_-$ by equation (1.4),  Hamilton's equations are of the form
$$
\frac{d}{dt}X = [((\lambda Y + X)^{k-1}\lambda^{-j})_+,(\lambda Y + X)] 
\tag{2.5}
$$
 where the subscript + denotes taking the polynomial part of the Laurent
polynomial $(\lambda Y +
X)^{k-1} \lambda^{-j}$.  To realize the equation (2.3) as one of  AKS -
type  we must find a
Hamiltonian in the AKS ring of spectral invariants  which gives the
equation (2.3).  Let the
$\lambda^i$ coefficient of
$(\lambda Y + X)^{k-1}$ be denoted by $A_i$.  Let
$$
q_j(\lambda)  = ((\lambda Y + X)^{k-1}\lambda^{-j})_+  \tag{2.6}
$$
and notice that
$$
q_j(0) = A_{k-1-j}  \tag{2.7}
$$ 
so in particular 
$$
\sum_{j=0}^{k-1} q_{k-j-1}(0)\mu ^j = (\mu Y + X)^{k-1}. \tag{2.8}
$$ 
Thus if we take 
$$
 H(\lambda, \mu, k) = \sum^{k-1}_{j=0} S_{k,k-j-1} \mu^j, \tag{2.9}
$$
we get a Lax pair which when evaluated at
$\lambda =0$ gives the Mischenko-Fomenko equation (2.3).

The proof for a more general Lie algebra $\frak{g}$ goes through {\it
mutatis mutandis} 
with $\sigma_k$ replaced by a general $Ad^*$- invariant function $\Phi$ and
$S_{k,j}$  replaced by
$\Phi (X(\lambda))_j$.

\bigskip
\noindent{\sectionfont 3  Spectral Darboux Coordinates}
\medskip

   We begin by recalling some of the results of {\bf [AHH2]}.
For the present, we continue with $\frak{g} = \frak{sl}(r)$, $G =
SL(r)$, and the pairing given by (2.1).  Given $\NN_0(\l)$ of the form (1.5) 
the coadjoint orbit through $\NN_0(\l)$ is given by
$$
\OO_{\NN_0} =\left\{\sum_{i=1}^n 
{g_iN_ig_i^{-1} \over (\l - \a_i)}\quad |\ g_i \in G \right\}  \tag{3.1}
$$
which may be identified with a product of $n$ coadjoint orbits in
$\frak{sl}(r)^*$).
For fixed $Y \in \frak{g}$ we consider shifted AKS Hamiltonians of the type
$$
\phi(\mu) = \Phi(Y + \mu),  \tag{3.2}
$$
where $\mu \in \OO_{\NN_0}$ and $\Phi$ is an element of the ring 
$I(\wt{\frak{g}}^*)$ of $Ad^*$ invariant functions on $\wt{\frak{g}}^*$.  
Hamilton's equations for
such functions are given by 
$$
{d\NN(\l)\over dt} = [d\Phi(\NN(\l))_+,\NN(\l)]  \tag{3.3}
$$
where $\NN(\l)$ has the form $Y + \mu$  and the $+$
subscript denotes projection to $\wt{\frak{g}}^+$.

   Setting 
$$
\LL(\l) := a(\l)\NN(\l)   \tag{3.4}
$$
with 
$$
a(\l) = \prod^n_{i = 1}(\l - \alpha_i), \tag{3.5}
$$ 
it is evident that the curve
$$
\SS_0 = \{(\l,z)\ |\ \det(\LL(\l) - zI) = 0\}   \tag{3.6}
$$
is invariant under the flows.  By the method of Krichever (see e.g.
{\bf [KN], [Du], [AHH1]}) the eigenspaces of $\LL^T(\l)$  define a line bundle
$E^*$ over  a suitable compactification $\SS$ of $\SS_0$  and this bundle moves
linearly in a component of $Pic(\SS)$ as $\LL(\l)$ evolves by (3.3).

    This geometric description of the linearization suggests the
following prescription for computing Darboux coordinates on $\OO_{\NN_0}$. 
First, let $\KK(\l,z)$ be the classical adjoint matrix of $\LL(\l) - zI$; i.e.
the matrix of cofactors transposed.  The rows of  $\KK(\l,z)$ give
sections of the eigenvector line bundle $E^*$ (over $\SS_0$).  Taking the
section of the dual
bundle $E$ obtained by restriction of a fixed section $(\l,z) \mapsto
((\l,z), V_0)$ of the 
trivial bundle $\SS_0 \times \Bbb C^r$, the zeros $\{(\l_{\mu},z_{\mu})\ :\
\mu = 1, ... d\}$ of
$\KK(\l,z)V_0$ give the associated  (spectral) divisor which, generically
is of degree $g + r -
1$.  If $V_0$ is not orthogonal to any of the eigenvectors of $\LL(\l)^T$
over $\l = \infty$, this
divisor is given by the $d = g+r-1$ finite solutions of the polynomial
equations 
$$
\KK(\l,z)V_0 = 0.  \tag{3.7}
$$
However, one must be careful if
$V_0$ is an eigenvector of the shift matrix $Y$. It then follows that $r-1$ of
the points in the divisor lie over $\l = \infty$.   Now let
$$
\z_{\mu} := \frac{z_{\mu}}{a(\l_{\mu})}   \tag{3.8}
$$ 
and consider 
$$
\{(\l_{\mu},\z_{\mu})\ :\ \mu = 1, ... d\}   \tag{3.9}
$$
 to be functions on the orbit $\OO_{\NN_0}$.  A
straightforward  computation {\bf [AHH2]}  shows that these functions
satisfy canonical relations, i.e. 
$$ 
\align
\{\l_{\mu},\l_{\nu}\} & = 0  \tag{3.10a} \\
\{\z_{\mu},\z_{\nu}\} & = 0  \tag{3.10b} \\
\{\l_{\mu},\z_{\nu}\} & = \delta_{\mu\nu}. \tag{3.10c}
\endalign
$$
This fact, together with some dimension counts yield the following theorem
(see{\bf [AHH2]}).

\proclaim {Theorem 3.1}\newline \noindent
a)  If $V_0$ is not orthogonal to any  eigenvector of $Y^T$, $d = g+r-1$
and the functions
$\{(\l_{\mu},\z_{\mu})\}$ give Darboux coordinates on a dense open set of
$\OO_{\NN_0(\l)}.$
\newline \noindent
b)  If $V_0 =  (1,0,0,...,0)^T$ and is an eigenvector of $Y$, then the functions
$\{\l_{\mu},\z_{\mu}\}$ together with 
$$
\z_i =(L_0)_{ii},\ \  \l_i = ln(L_1)_{i1}  \tag{3.11}
$$
$i = 1, ... n-1,$ give Darboux coordinates.  Here $L_0$ and
$L_1$ are determined by 
$$\LL(\l) = a(\l)Y + L_0\l^{n-1} + L_1\l^{n-2} + ...\ .  \tag{3.12}  
$$
\endproclaim

   We remark that part b) of this theorem may be stated for more general
$V_0$ but 
 a change of basis may always be used to transform it to this form.

    To illustrate how this result applies to semisimple Lie algebras, we
consider  the case when
$\NN_0(\l)$ has the form (1.4) with  $X$ in $\frak{sl}(2),
\frak{sl}(3)$ or $\frak{so}(3).$  Here 
$a(\l) = \l$ so
$$\z_i = z_i/\l_i.   \tag{3.13}$$
  Note that although the
theorem is proved for $\frak{sl}(r,\Bbb C)$ or $\frak{sl}(r,\Bbb R)$, it
may be applied to
subalgebras defined as fixed points of a finite group of automorphisms,
provided the resulting
coordinates are invariant under this  group.
\smallskip \noindent
(a.1)\ Take $\frak{g}=\frak{sl}(2,\bold R)$.
The dimension of a generic orbit is
dim~$\OO_{\NN_0} =2$.
We parametrize $\NN_0(\l)$ as follows:
$$
\NN_0(\l) = { N_1\over \l}  :=  {1\over \l}\pmatrix -a & r
              \\ u & a \endpmatrix,  \tag{3.14}
$$
and choose
$$
Y := \pmatrix 1 & 0 \\ 0 & -1 \endpmatrix, \quad V_0 = \pmatrix 1\\0
\endpmatrix.
\tag{3.15}
$$
The characteristic equation is then
$$
\text{det } (\LL(\l) - z \Bbb I_r)= z^2 - \l^2 + 2\l a - a^2 - u r =0. 
\tag{3.16}
$$
In this case,  $V_0$ is an eigenvector of $Y$ and the genus of the spectral 
curve is $g=0$, so there are no $\{\l_\mu, \z_\mu\}$'s. The single pair
of spectral Darboux coordinates is thus
$$
q_2 = \text{ln }u, \quad P_2 = a.  \tag{3.17}
$$
It is easily verified that, relative to the Lie Poisson structure,
they satisfy
$$
 \quad \{q_2, P_2\} =1. \tag{3.18}
$$
\smallskip \noindent
(a.2)\  Consider the same orbit as in (a.1), but choose
$$
Y := \pmatrix 0 & 1 \\ 1 & 0 \endpmatrix.  \tag{3.19}
$$
In this case, $V_0$  is not an eigenvector of $Y$. The genus is still
0 but the line bundle does have a finite divisor point, giving the
Darboux  coordinate pair
$$
\l_1 = -u, \quad \z_1 =-{a \over u}. \tag{3.20}
$$
These are verified to also satisfy
$$
 \{\l_1, \z_1\}=1. \tag{3.21}
$$
\smallskip \noindent
(b.1)\ Take $\frak{g}=\frak{sl}(3, \bold R)$. The
dimension of a generic orbit  is dim $\OO_{\NN_0} =6$. We
parametrize $\NN_0(\l)$ as:
$$
\NN_0(\l) :=  {1\over \l} \pmatrix -a -b & r & s \\
               u & a & e \\
               v & f & b \endpmatrix, \tag{3.22}
$$
and choose
$$
Y = \pmatrix 0 & 0&0\\ 0 & 1 & 0 \\ 0 & 0 & -1  \endpmatrix,
 \quad V_0 = \pmatrix 1\\0\\0\endpmatrix.  \tag{3.23}
$$
Again, $V_0$ is an eigenvector of $Y$, but the spectral curve has genus
$g=1$, and is realized as a $3$--fold branched cover of $\Bbb P^1$. We
therefore find one Darboux coordinate pair $(\l_1, \z_1)$, corresponding to a
finite zero of the eigenvector components, plus two further pairs,
$(q_2, P_2, q_3, P_3)$, corresponding to zeros over $\l=\infty$:
$$
\align
 \l_1 = {1\over 2} \left(b-a-{e v\over u} + {u f \over v}\right),&\quad
\z_1 = {u v  a + u v  b - e v^2 - f u^2 \over -u v  a + u v  b - e v ^2 +
 f u^2}
\\
q_2= \text{ln }u, \quad q_3= \text{ln }v,&\quad P_2 = a,\quad P_3 = b.
\tag{3.24}
\endalign
$$
Again, it is easily verified directly that these form a Darboux system,
with nonvanishing Lie Poisson brackets
$$
\{\l_1, \z_1\}=1,\quad \{q_2, P_2\} =1,\quad \{q_3, P_3\}=1.  \tag{3.25}
$$
\smallskip \noindent
(b.2)\   Consider the same orbit as in (b.1), but take 
$$
Y = \pmatrix 0 & 1 &0\\ 1 & 0 & 1 \\ 0 & 1 & 0  \endpmatrix,
 \quad V_0 = \pmatrix 1\\0\\0\endpmatrix.  \tag{3.26}
$$
The genus of the curve is still 1, but the number of finite divisor points
is now 3. 
The divisor coordinates may be obtained by solving the pair of equations
$$
z^2 + z(v-a-b) + \l (u-e) +ab - ef +fu -av =0 \tag{3.27a}
$$
$$
\l z +uz +(v-b)\l +ev -bu = 0, \tag{3.27b}
$$
which reduces to a cubic equation for $z$, with generically distinct roots
$(z_1,z_2,z_3)$.  Setting 
$$
\l_i = \frac{z_i^2 +(v-a-b)z_i +ab-av+fu-fe}{u-e},\ \  \z_i = \frac {z_i}{\l_i},
\ \ i=1,2,3 \tag{3.28}
$$ gives Darboux coordinates.
\smallskip \noindent
(c)\ Take $\frak{g}=\frak{so}(3, \bold R)$.   (The proof of Theorem 3.1 was
given only in the
case $\frak{g}= \frak{sl}(r)$, but we include this example to illustrate
that it can be extended
to a more general setting.  The explicit results can be readily checked by
direct computation.)  
The dimension of a generic orbit  is dim
$\OO_{\NN_0} = 2$. We parametrize $\NN_0(\l)$ as:
$$
\NN_0(\l) :=  {1\over \l} \pmatrix 0 & r & s \\
               -r & 0 & e \\
               -s & -e & 0 \endpmatrix. \tag{3.29}
$$
Using $Y$ and $V_0$ as in (3.23) we get a spectral curve of genus 1 and one
pair of divisor coordinates given by making the appropriate restrictions of
$(\l_1,\z_1)$ in (3.24).  Namely,
$$
\l_1 = -\frac{1}{2}\frac{e(s^2 +r^2)}{sr}, \ \ \z_1 = \frac{s^2-r^2}{r^2 + s^2}.
\tag{3.30}
$$
A direct computation shows that for the Lie Poisson structure on
$\frak{so}(3, \bold
R)$ one has 
$$\{\l_1,\z_1\} = 2.  \tag{3.31}$$
Thus, one adjusts $\z_1$ by a factor of
$\frac{1}{2}$ to get a canonical pair.  (This is a remnant of the fact that
we used the pairing
(1.2) to identify $\frak{g}$ with $\frak{g}^*$, while here the appropriate
pairing is one half of
the pairing (1.2).)

   More generally, it is interesting to  compute the Darboux coordinates one
gets on these orbits using 
$$
Y = \pmatrix \alpha & 0 &0\\ 0 & \beta & 0 \\ 0 & 0 & \gamma  \endpmatrix,
  \tag{3.32}
$$
with distinct $\alpha , \beta, \gamma$.  Rigid body motion in 3 dimensions
can be written in
the Lax pair form 
$$\frac{d\LL}{dt} = [\Cal A, \L]  \tag{3.33}$$
where
$$
\LL = \l Y + \Lambda  \tag{3.34}
$$
with the diagonal matrix
$$
Y = \diag(\alpha, \beta, \gamma)  \tag{3.35}
$$
given by the inverses of the principal moments of inertia
$$
\alpha = \frac{1}{I_1}, \ \ \beta = \frac{1}{I_2}, \ \ \gamma =
\frac{1}{I_3}  \tag{3.36}
$$
and $\Lambda \in \frak{o}(3)$ the components of the angular momentum vector
 relative to the principal axes of inertia:
$$
\Lambda  =    \pmatrix 0 & L_3 & -L_2 \\
               -L_3 & 0 & L_1 \\
               L_2 & -L_1 & 0 \endpmatrix, \tag{3.37}
$$
while 
$$
\Cal A = \frac{1}{\l}\Lambda^2.   \tag{3.38}
$$
Thus, identifying
$$ 
e = L_1, \ \ s = - L_2, \ \  r = L_3,  \tag{3.39}
$$
we have
$$\LL = \l (Y +N_0(\l)).  \tag{3.40}
$$
The spectral curve for this case is given by
$$
\align
z^3 - z^2\l(\a +\b +\g) +z\l^2 (\a\b  + \a\g + \b\g ) -   \l^3(\a\b\g) - &\\
\l(\a e^2 +\g r^2 +\b s^2) + z(r^2 + s^2 +e^2) & = 0.  \tag{3.41}
\endalign
$$
The coefficient of the linear terms in $z$ and $\l$ are just the square of
the angular momentum
vector
$$ |\bold{L}|^2 = e^2 +s^2 +r^2 = L_1^2 + L_2^2 + L_3^2
\tag{3.42}
$$
and the Hamiltonian for the Euler top
$$
H=\frac{1}{2}(\a e^2 + \b s^2 + \g r^2) =
\frac{1}{2}\left[\frac{L_1^2}{I_1}+\frac{L_2^2}{I_2}+\frac{L_3^2}{I_3}\right
].  \tag{3.43}
$$
Choosing the vector
$$ 
V_0 =(1,0,0)^T \tag{3.44}
$$
and noting that the pairing needed to identify $\frak{o}(3)$ with its dual is 
$$
<X,Y> = -\frac{1}{2}tr(XY),  \tag{3.45}
$$
the Darboux coordinates provided by theorem (3.1) are:
$$
\frac{1}{2}\l_1 =  \frac{e(r^2 +s^2)}{2rs(\b - \g)}, \ \ \z_1 = \frac{\b
s^2 + \g r^2}{(r^2
+s^2)}.  \tag{3.46}
$$
(Note that the factor ${1\over 2}$ in $\l$ is again required because we are
not dealing with the
full
$\wt{\frak{sl}}(3)$ algebra, but the subalgebra consisting of fixed points
under the involution
$X(\l)\ra -X(-\l)^T$. Note also that, although $V_0$ is an eigenvector of
$Y$, in this case the
orbits are $2$--dimensional, and part (b) of theorem 3.1 provides no
additional coordinates.)

  On the level sets given by fixing the values of the invariants $H$ and
$|\bold{L}|^2$ as
$$
H=E, \qquad |\bold{L}|^2 = \ell^2,  \tag{3.47}
$$
the coordinate function $\z_1$ is just
$$
\z_1 = {2E -\a L_1^2 \over \ell^2 -L_1^2}. \tag{3.48}
$$

   For purposes of integration, it is convenient to parametrize the spectral
curve in a more
standard form that makes it evident it is elliptic. In terms of $(\l,
\z={z\over\l})$ we may express
eq. (3.41) as:
$$
\l^2 a(\z) + \z \ell^2 -2E=0,  \tag{3.49}
$$
where
$$
a(\z):= (\z-\a)(\z-\b)(\z-\g).  \tag{3.50}
$$
Then, defining the meromorphic function
$$
y := \frac{2E - \ell^2\z}{\l}   \tag{3.51}
$$
the expression (3.47) is equivalent to
$$
y^2 = (2E- \ell^2\z )a(\z).  \tag{3.52}
$$
We may therefore interpret the spectral curve $\SS_0$ as the Riemann
surface of the function $y$.

   The flow may then be computed through the Liouville method using the Darboux
coordinates $(\l_1/2, \ \z_1)$. Restricting the $1$--form
$$
\theta = - {\l_1\over 2} d\z_1  \tag{3.53}
$$ 
to the Lagrangian leaf defined by eq. (3.47) and integrating, the Liouville
generating function is
$$
S(\z_1, E, \ell) = -{1\over 2}\int_{(\z_o, \l_0)}^{(\z_1, \l_1)}\l d\z
= -{1\over 2} \int_{(\z_o, \l_0)}^{(\z_1, \l_1)} {(2E-\ell^2 \z)\over y}
d\z  \tag{3.54}
$$
(where we have used the fact that the point $(\l_1, z_1=\z_1 \l_1)$ lies on
the spectral curve).
Differentiating with respect to $E$ then gives the conjugate coordinate to
$H$, which evolves
linearly in time
$$
Q:= {\di S \over \di E}
 = -{1\over 2}\int_{(\z_0, \l_0)}^{(\z_1, \l_1)} {d\z\over\sqrt{(2E -
\ell^2\z)(a(\z))}} =t-t_0, 
\tag{3.55}
$$ 
where a base point $(\z_0, \l_0) = (\z_1(t_0), \l_1(t_0))$ has been chosen. As
usual, the elliptic
integral appearing in eq. (3.55) may be inverted to determine $\z_1(t)$,
and hence $(L_1(t), \
L_2(t), \ L_3(t))$ in terms of Jacobi elliptic  functions {\bf[W]}.

\bigskip
\noindent{\sectionfont 4  Duality}
\medskip

We now extend this method to the case of spectral Darboux coordinates on
rank two orbits in
$\frak{so}(n)^*$.  First we  describe these  using $n \times n$ matrices
and then,  using
duality and  a judicious choice of $V_0$, we  show that the method can be
used to deduce   the
hyperelliptic cooordinates used to integrate  Euler flow on these orbits as
in {\bf[Mo]}.

   The Euler flow on $\frak{so}(n)^*$ is generated by a quadratic spectral
invariant of the matrix
$$
M(z) = A + \frac{1}{z}L,\ \  L \in \frak{so}(n),  \tag{4.1}
$$
 where $A$ is
a diagonal $n \times n$ matrix.  If  we take $L$ to have rank two, we may
write it in the form 
$$
L = X \wedge Y = XY^T - YX^T  \tag{4.2}
$$
for some $X, Y \in \Bbb R^n$.   This provides a map from $\Bbb R^{2n}$ to
the space of rank two $n
\times n$ skew matrices.  It is easily checked that this is a Poisson map
into the one
parameter family of rank two coadjoint orbits in  $\frak{so}(n)$.  The only
nonvanishing coadjoint
invariant is
$$
tr (L^2) = 2((X\cdot Y)^2 -|X|^2|Y|^2)  \tag{4.3}
$$ 
and the fiber of this map is given by the orbits of the
$\frak{sl}(2)$ action on
$\Bbb R^{2n}$ given by
$$
g\cdot (X,Y) = (aX+bY,cX+dY) \tag{4.4}
$$  where
$$ g = \pmatrix a & b\\ c & d 
\endpmatrix .\tag{4.5}$$   Thus the coadjoint  orbit is $2n - 4$
dimensional. In the following we
will describe the divisor coordinates for $M(z)$ as functions  in the variables
$(X,Y)$ that project under the $\frak{sl}(2)$ quotient to the orbits of the
form (4.2).
 
If we let
$\widetilde{M(z) -\l I}$ denote the transposed matrix of cofactors of
$M(z) - \l I$ and  choose $V_0 \in \Bbb R^n$  then, as above, the divisor
Darboux coordinates
are given by the zeros of $(\widetilde{M(z) -\l I})V_0$.  
 Note that
$$
\widetilde{M(z) -\l I} = (\widetilde{A - \l I})(\widetilde{I +
\frac{1}{z}L(A-\l I)^{-1}})
\tag{4.6}
$$
 so
$$
\widetilde{(M(z) -\l I)}V_0 = 0 \tag{4.7}
$$ 
if and only if
 $$
\widetilde{(I +\frac{1}{z}L(A- \l I)^{-1})}V_0 = 0. \tag{4.8}
$$ On these orbits, we can compute
$\widetilde{I +\frac{1}{z}L(A-
\l I)^{-1}}$ explicitly.  First, by direct computation, one sees that 
$$ 
(L(A- \l I)^{-1})^3 = \g L(A- \l I)^{-1}  \tag{4.9}
$$
where 
$$
 \g = Q(X,Y)^2 - Q(X,X)Q(Y,Y) \tag{4.10}
$$ 
with 
$$
Q(U,V) = U^T(A - \l I)^{-1}V. \tag{4.11}
$$ 
 Thus, the minimal
polynomial for $I +\frac{1}{z}L(A- \l I)^{-1}$ is given by
$$
u^3 -3u^2 +(3- \frac{\g}{z^2})u = (1 - \frac{\g}{z^2}).  \tag{4.12}
$$
From (4.12), it follows that 
$$
\widetilde{(I +\frac{1}{z}L(A- \l I)^{-1})} = \frac{1}{z^2}(L(A- \l
I)^{-1})^2 -\frac{1}{z}L(A- \l
I)^{-1} +(1 - \frac{\g}{z^2})I .  \tag{4.13}
$$
Applying this matrix to $V_0$, we obtain
$$
\align &\widetilde{(I +\frac{1}{z}L(A- \l I)^{-1})}V_0 =  \tag{4.14}\\
&(1 - \frac{\g}{z^2})V_0  
 +[\frac{1}{z^2}(Q(X,Y)Q(Y,V_0) -Q(Y,Y)Q(X,V_0)) - \frac{1}{z}Q(Y,V_0)]X \\
&+[\frac{1}{z^2}(Q(X,Y)Q(X,V_0) - Q(X,X)Q(Y,V_0)) + \frac{1}{z}Q(X,V_0)]Y.  
\endalign
$$

   For fixed $V_0$ and generic values of $X$ and $Y$, these three terms 
are linearly
independent, so the above expression vanishes if
$$ \align   \g &= z^2, \tag{4.15a}\\
Q(X,Y)Q(Y,V_0)& -Q(Y,Y)Q(X,V_0) -zQ(Y,V_0) = 0,\tag{4.15b}\\
Q(X,Y)Q(X,V_0)& - Q(X,X)Q(Y,V_0) +zQ(X,V_0) = 0.\tag{4.15c}
\endalign 
$$
The first of these follows from the last two, and these two are equivalent to
$$
\align
 Q(X,X)Q(Y,V_0)^2 - 2Q(X,Y)Q(X,V_0)Q(Y,V_0) + & Q(Y,Y)Q(X,V_0)^2 =0
\tag{4.16a}\\
z = Q(X,Y) - Q(Y,Y)\frac{Q(X,V_0)}{Q(Y,V_0)}&. \tag{4.16b}
\endalign
$$
Equation (4.16a) determines the $\lambda_{\mu}$'s as roots of a polynomial,  and
substituting these into (4.16b) gives the $z_{\mu}$'s.  For example, if we take
$V_0 =(1,0,0, \dots 0)^T$, then (4.16a) is just 
$$
0=Q(X,X)y_1^2 - 2Q(X,Y)x_1y_1 +Q(Y,Y)x_1^2  \tag{4.17}
$$
 which, when multiplied by $\prod_{j=2}^n (\l -\a_j)$, gives a polynomial
equation of degree $n-2$
in
$\l$.  Notice that (4.16a,b) are invariant under the $\frak{sl}(2)$ action
on $\Bbb R^{2n}$, so the
$n -2$ pairs
$(\l_{\mu},z_{\mu})$ reduce to the quotient, which is identified with the
coadjoint orbit through
L. 

   We now turn to the dual description {\bf [AHH3]} of divisor coordinates
for these rank two
coadjoint orbits in $\frak{so}(n)^*$.  Instead of the map $(X,Y)
\rightarrow M(z)$ considered above
we take the map  $\Bbb R^{2n} \rightarrow \tilde{\frak{sl}}(2,\bold R)_+^*$
given by
$$
(X,Y) \rightarrow N(\l) = \sum_{i=1}^n \frac{1}{\l - \alpha_i}\pmatrix
-x_iy_i & y_i^2 \\ -x_i^2 &
x_iy_i
\endpmatrix . \tag{4.18}
$$
This is a Poisson map whose generic fibers are the $2^n$ points in an orbit
of the group that sends
$(x_i,y_i)$ to $(\pm x_i,\pm y_i)$.  We can identify the coadjoint orbit
through $M(z) \in
\tilde{\frak{so}}(n)_+^*$ with the  symplectic quotient by the
$\frak{sl}(2)$ action on the
coadjoint orbit  through $N(\l) \in \tilde{\frak{sl}}(2)_+^*$. Indeed, the
$\frak{sl}(2)$ moment 
map is given by 
$$
N_0 =
\sum_{i=1}^n \pmatrix -x_iy_i & y_i^2 \\ -x_i^2 & x_iy_i
\endpmatrix \tag{4.19}
$$  
so, for instance, the orbit in   $\frak{so}(n)$ with $\tr (L^2) = -2$ can
be realized as
the quotient of the set 
$$
\{(X,Y) \in \Bbb R^{2n} |\ \  |X| =|Y| = 1, X \cdot Y =0\} \tag{4.20}
$$
 by the action
of
$\frak{so}(2) \subset \frak{sl}(2)$.  

    To get divisor coordinates on the orbit through $N_0(\l)$ it is
standard to choose $V_0 =
(1,0)$, giving the equations
$$
Q(X,X) = 0,\ \ \   z = Q(X,Y).   \tag{4.21}
$$
The first of these equations gives hyperellipsoidal coordinates on the
sphere.  However, these
coordinates are not suited for studying the rigid body equations since they
are not invariant 
under the $\frak{so}(2)$ action and hence do not reduce to the
$\frak{so}(n)$ orbit.  To
circumvent this problem we must choose $V_0$ so that the resulting divisor
coordinates are
$\frak{so}(2)$ invariant.  One such choice is
$$
V_0 = (y_1,x_1), \tag{4.22}
$$
which leads to the equation (4.17).   Since equation (3.7) for this choice
of $V_0$ is invariant
under the $\frak{so}(2)$ action
$$
(X,Y) \mapsto (cos\theta X +sin\theta Y, -sin \theta X + cos\theta Y),
\tag{4.23}
$$
these spectral divisor coordinates project to the reduced space.   This is
a nonstandard choice of
$V_0$ since it is dependent on the point in the coadjoint orbit.  However 
it still defines a
section of the dual eigenvector line bundle over the spectral curve, and 
its  zeros still give
Darboux coordinates on the coadjoint orbit.  

    Another way to assure invariance under the $\frak{so}(2)$
action is to use the fixed complex vector 
$$
V_0 =(1,i)  \tag{4.24}
$$
which leads to the equations
$$
Q(X+iY,X+iY) = 0, \ \ \  z = Q(X+iY,Y). \tag{4.25}
$$
Since the action of an element of $\frak{so}(2)$ on $V_0$ simply amounts to
multiplication of
$V_0$ by a phase $exp(i\theta )$, the resulting divisor coordinates are
again $\frak{so}(2)$
invariant, and thus reduce to the $\frak{so}(n)$ orbit.  Finally, we remark
that equations (4.25)
can be found in the $n \times n$ setting by choosing $V_0= X + iY$ in 
equations (4.16a,b).  This
is again a ``dynamical" choice of section of $E$ (i.e., it depends on the
point in the orbit
determined by the pair $(X,Y)$), just as the choice $V_0 = (y_1,x_1)$ was
in the $2 \times 2$
setting.  But, by equivalence of these dual formulations, it again provides
a valid spectral
Darboux system on the orbit.

\newpage
\centerline{\smc References}
\bigskip
{\smaller{
\item{ \bf [Ad]} Adler, M., ``On a Trace Functional
for Formal Pseudo-Differential Operators and the Symplectic
Structure of the Korteweg- de Vries Equation'', {\it Invent. Math.}
{\bf 50}, 219-248 (1979). 
\item{\bf [AHH1]} Adams, M.R., Harnad, J. and Hurtubise, J., ``Isospectral
Hamiltonian Flows in Finite and Infinite Dimensions II.
Integration of Flows'',
 {\it Commun\. Math\. Phys\.} {\bf 134}, 555--585 (1990).
\item{\bf [AHH2]}  Adams, M.R., Harnad,~J. and  Hurtubise,~J., ``Darboux
Coordinates and Liouville-Arnold Integration  in Loop Algebras,''
  {\it Commun\.  Math\.  Phys\.}{\bf 155}, 385-413 (1993).
\item{\bf[AHH3]} Adams, M.R., Harnad, J. and Hurtubise, J.,
  ``Dual Moment Maps
into Loop Algebras", {\it Lett. Math. Phys.} {\bf 20}, 299-308  (1990).
\item{\bf[AHP]} Adams, M.R., Harnad, J. and Previato, E.,
``Isospectral Hamiltonian Flows in Finite and Infinite
Dimensions I. Generalised Moser Systems and Moment Maps
into Loop
Algebras'', {\it Commun. Math. Phys.} {\bf 117}, 451-500
(1988).
\item {\bf [AM]} Abraham, R., Marsden, J.  ``Foundations of Mechanics", 2nd ed.
Benjamin/Cummings, Reading, Mass. (1978).
\item{\bf[Ar]}  Arnol\'d,  V.I., {\it Mathematical Methods of Classical
Mechanics}, 
Springer-Verlag, New York, 1978.
 \item{ \bf[AvM]} Adler, M. and
van Moerbeke, P., ``Completely
Integrable Systems, Euclidean Lie Algebras, and Curves'',
{\it Adv. Math.}
{\bf 38}, 267-317 (1980).
\item{\bf [Du]} Dubrovin, B.A., ``Theta Functions and Nonlinear Equations'',
 {\it Russ\. Math\. Surv\.} {\bf 36}, 11--92 (1981).
\item {\bf [GS]} Guillemin, V. and Sternberg, S.  ``The Moment Map and
Collective Motion", {\it
Annals of Physics} {\bf 127}, 220--253 (1980).
\item {\bf [K]} Kostant, B., ``The Solution to a Generalized Toda Lattice and
Representation Theory'', {\it Adv. Math}, {\bf 34} (1979), 195-338.
\item{\bf [KN]} Krichever, I.M. and  Novikov, S.P.,  ``Holomorphic Bundles over
Algebraic Curves and Nonlinear Equations'', {\it Russ\. Math.~ Surveys}
{\bf 32}, 53--79 (1980).
\item {\bf[MF]} Mischenko, A.S., Fomenko, A.T., ``On the
integration of the Euler equations on semisimple Lie algebras",
{\it Sov. Math. Dokl.} {\bf 17}, 1591-1593 (1976)
\item{ \bf[Mo]}  Moser, J.,  ``Geometry of Quadrics and Spectral Theory",{\it
The Chern Symposium, Berkeley, June 1979}, 147-188, Springer, New York,
(1980).
\item{\bf[R]}  Ratiu, T.,  ``The Lie
algebraic interpretation of the complete integrability of the \newline
Rosochatius system", {\it Mathematical Methods in Hydrodynamics 
 and
Integrability in  Dynamical Systems, AIP Conference Proc.} {\bf 88},
La Jolla (1981),  ``The motion of the free n-dimensional rigid
body", {\it Ind. Univ. Math. J.} {\bf 29}, 609-629, (1980).
\item { \bf[S]} Symes, W., ``Systems of Toda Type, Inverse
Spectral Problems and Representation Theory'', {\it Invent. Math.},
{\bf 13 - 59} (1980).
\item{ \bf[W]}  Whittaker, E.T.,  `A Treatise on the Analytical Dynamics of
Particles and
Rigid Bodies'', 4th ed. Cambridge University Press, Cambridge (1959).

\medskip}}
\vfil\eject

\end